\documentclass[aps,prd,preprint,groupedaddress,showpacs]{revtex4}

\usepackage{graphicx}
\usepackage{bm}

\newcommand{\grad}{{\rm grad}}
\newcommand{\divv}{{\rm div}}

\begin{document}

\title{Plasma cutoff and enhancement of
radiative transitions in dense stellar matter}

\author{P.~S.~Shternin}\email{pshternin@gmail.com}
\author{D.~G.~Yakovlev}
\affiliation{Ioffe Institute, Polytekhnicheskaya Street 26,
Saint-Petersburg 194021, Russia}

\date{\today}

\begin{abstract}
We study plasma effects on radiative transitions (e.g., decay of
excited states of atoms or atomic nuclei) in a dense plasma at the
transition frequencies $\omega \lesssim \omega_p$ (where
$\omega_p$ is the electron plasma frequency). The decay goes
through four channels -- the emission of real transverse and
longitudinal plasmons as well as the emission of virtual
transverse and longitudinal plasmons with subsequent absorption of
such plasmons by the plasma. The emission of real plasmons dies
out at $\omega \leq \omega_p$, but the processes with virtual
plasmons strongly enhance the radiative decay.
Applications of these results to radiative processes
in white dwarf cores and neutron star envelopes are discussed.
\end{abstract}


\maketitle

\section{Introduction}
\label{S:intro}

Radiative processes in stars are very important. First of all, they
determine heat transport in radiative zones of the stars
\cite{cg68}, as well as the radiative transfer and structure of
stellar atmospheres together with the formation of spectra of
stellar radiation \cite{chandra60,sobolev69}. In ordinary stars (at
the main sequence or around) typical radiation frequencies are much
higher than the electron plasma frequency $\omega_p$ of stellar
matter. As a result, plasma effects do not affect strongly radiative
processes.

However, in a dense matter of compact stars (white dwarfs and
neutron stars) the plasma frequency can be higher or comparable to
characteristic radiation transition frequencies $\omega$, and the
plasma effects cannot be ignored. For instance, in a strongly
degenerate nonrelativistic electron gas at a density of
$10^3$~g~cm$^{-3}$ (that is typical for degenerate cores of white
dwarfs and outer envelopes of neutron stars) one has $\hbar \omega_p
\approx 0.6$ keV. In this example the plasma effects can easily
affect radiative transitions in atoms and ions. In the inner crust
of a neutron star at a density of $10^{12}$ g~cm$^{-3}$, where the
degenerate electrons are ultrarelativistic, the plasma frequency
becomes very large, $\hbar \omega_p \sim 3$ MeV (depending on the
composition of the crust; e.g., Ref.~\cite{hpyBOOK}; also see
Sec.~\ref{S:dissc}). This is large enough to influence radiative
transitions in atomic nuclei.

The importance of plasma effects for radiative processes in dense
stellar matter has been mentioned in the literature (e.g., Ref.\
\cite{by76}). In particular, the plasma effects on the radiative
thermal conductivity have been studied in Refs.\ \cite{ao79,kvh94}
but these studies are not fully complete (Sec.\ \ref{S:dissc}). For
another example, consider an emitter (an ion or atomic nucleus) in
an excited state with the transition frequency $\omega$ to the
ground state that is lower than $\omega_p$. What will happen with
this emitter taking into account that radiative transitions with the
emission of any electromagnetic quanta are now forbidden? Will it
live at the excited level forever?

These questions can be answered using the available theory of
electromagnetic transitions in a plasma. The plasma impact on
electromagnetic transitions in an non-relativistic laboratory plasma
and a rarefied non-relativistic cosmic plasma has been studied for a
long time (e.g., Refs.\ \cite{wm98,OiKl84}). The plasma effects can
modify the emission of electromagnetic quanta \cite{OiKl84}.
Moreover, collective plasma processes open another electromagnetic
transition channel -- the emission of virtual plasmons and
successive absorbtion of these plasmons by the plasma
\cite{Klim84,wm98}. The most pronounced of these effects seems to be
collisional broadening of energy levels of atoms and ions and
associated broadening of spectral lines. It can be important in
cosmic and laboratory plasmas \cite{Klim84}.

To the best of our knowledge, the theory of radiative transitions
in a plasma (e.g., Refs.\ \cite{Klim84,wm98}) has been correctly
applied only to study non-degenerate laboratory and cosmic
plasmas. In this paper we investigate the radiative transitions in
a dense degenerate relativistic electron gas, particularly at
$\omega<\omega_p$. The paper is organized as follows. In Sec.\
\ref{S:four} we outline the formalism for calculating
electromagnetic transitions rates in a dense plasma. It is similar
to the formalism of stopping power for a charged particle moving
in a plasma \cite{abr84eng}. In Sec.~\ref{S:diel_funct} we outline
the main properties of a degenerate electron plasma. Section
\ref{S:el_dip} is devoted to the radiative decay in the plasma for
those transitions that are allowed in the electric dipole
approximation. In Sec.\ \ref{S:quadr} we address similar problem
for the electric quadrupole and magnetic dipole transitions. In
Sec.\ \ref{S:dissc} we discuss the main results and some
applications, particularly, for calculating radiative thermal
conductivity in white dwarf cores and neutron star envelopes and
for studying radiative decay of excited states of atomic nuclei
and kinetics of neutrons in the neutron star crust. We conclude in
Sec.\ \ref{S:concl}.


\section{Four radiative transition channels} \label{S:four}

Let us consider an external emitter (for instance, an atom or atomic
nucleus) immersed in a plasma.  The plasma is assumed to be uniform
and isotropic; it is characterized by the longitudinal and
transverse dielectric functions $\varepsilon_\mathrm{l}(\omega,k)$
and $\varepsilon_\mathrm{tr}(\omega,k)$, respectively. We are
interested in the transition rate $w_{i\to f}$ [s$^{-1}$] at zero
temperature ($T=0$) from an upper state $i$ to a lower state $f$
whose energy separation is $\hbar \omega$. The expression for
$w_{i\to f}$
can be written as \cite{Klim84}
\begin{equation}
\label{eq:wif}
    w_{i\to f} = -\frac{e^2}{\pi^2\hbar}\Im \int\limits_0^\infty
    {\rm d}k\int\limits_{(4\pi)} {\rm d}\Omega_{\bm{k}}
    \left[\frac{\left|\bm{j}_{fi}(\bm{k})\bm\cdot \bm{k}\right|^2}
    {\omega^2\varepsilon_\mathrm{l}(\omega,k)}+
    \frac{\left|\bm{j}_{fi}(\bm{k}) \bm \times \bm{k}\right|^2}
    {\omega^2\varepsilon_\mathrm{tr}(\omega,k)-k^2c^2}\right].
\end{equation}
Any elementary transition (characterized by the given energy loss
$\hbar \omega$) is accompanied by the transfer of an elementary
excitation with a wavevector $\bm{k}$ to a plasma coupled to
electromagnetic field. The integration is performed over all
allowed values of $\bm{k}$ (with $k=|\bm{k}|$, and
d$\Omega_{\bm{k}}$ being a solid angle element in the direction of
$\bm{k}$).  Furthermore,
\begin{equation}
     \bm{j}_{fi}(\bm{k})=\int {\rm d} V\, \bm{j}_{fi}(\bm{r})
     \exp(-i\bm{k}\bm{r})
\label{eq:jfi}
\end{equation}
is the Fourier transform of the local transition current
$\bm{j}_{fi}(\bm{r})$ \cite{landau4eng}. The latter current can be
calculated from relativistic theory with stationary (relativistic)
wave functions of the emitter in the $i$ and $f$ states. The
transition rate (\ref{eq:wif}) is cumulative. It includes
contributions of transition channels with different $\bm{k}$ and
different structures of plasma-electromagnetic field excitations.
Particularly, it intrinsically contains a sum over polarizations of
emitted plasmons (see below). It neglects the contribution of two-
and multiple-plasmon processes which is expected to be small in the
cumulative rate.

The dielectric functions (with spatial dispersion) in
Eq.~(\ref{eq:wif}) take into account the plasma effects on the
transition rate \cite{wm98,Klim84}. In particular,
Eq.~(\ref{eq:wif}) includes the contribution of direct interaction
of the emitter with plasma electrons.

The transition rate (\ref{eq:wif}) can be decomposed as (Table~\ref{tab1})
\begin{equation}
\label{eq:wif_tot}
      w_{i\to f}=w_{i\to f}^\mathrm{l}+w_{i\to f}^\mathrm{tr} =
      w_{i\to f}^\mathrm{Al}+w_{i\to f}^\mathrm{Bl}
      +w_{i\to f}^\mathrm{Atr}+w_{i\to f}^\mathrm{Btr}.
\end{equation}
Here, $w_{i\to f}^\mathrm{l}$ and $w_{i\to f}^\mathrm{tr}$
correspond to the longitudinal and transverse channels [the terms in
(\ref{eq:wif}) containing $\varepsilon_\mathrm{l}(\omega,k)$ and
$\varepsilon_\mathrm{tr}(\omega,k)$, respectively]. Each of these
terms, in turn, contains two contributions -- (A) the emission of a
real longitudinal (Al) or transverse (Atr) plasmon and (B) the
emission and absorption of a virtual longitudinal (Bl) or transverse
(Btr) plasmon.
Not all of the four channels can be opened at once (see
below).

\begin{table}[t]
\caption[]{Four transition channels
in plasma environment.}
\label{tab1}
\begin{center}
\begin{tabular}{c c c c  }
\hline
\hline
Channel&~Plasmon~&~Open at~&~Comment~ \\
\hline 
Atr  & real transverse &  $\omega_p<\omega$ &
Dominates at $\omega \gtrsim \omega_p$ \\
Al  & real longitudinal &  $0<\omega-\omega_p \lesssim \omega_p$ &
  \\
Btr  & virtual transverse &  any $\omega$ &
  \\
Bl  & virtual longitudinal & any $\omega$ &
Dominates at $\omega \lesssim \omega_p$ \\
\hline
\hline
\end{tabular}
\end{center}
\end{table}

The emission of real plasmons (channels Al and Atr) is allowed in
the presence of the poles in the denominators of Eq.~(\ref{eq:wif}),
that is at
\begin{equation}
\label{eq:poles}
    \varepsilon_\mathrm{l}(\omega,k)=0, \quad
    \omega^2\varepsilon_\mathrm{tr}(\omega,k)=k^2c^2.
\end{equation}
%
The roots of these equations give the plasmon dispersion relations
$k_\mathrm{l}(\omega)$ and $k_\mathrm{tr}(\omega)$ for
longitudinal and transverse plasmons, respectively. The emission
rates for real longitudinal and transverse plasmons are then given
by the standard expressions \cite{OiKl84}
\begin{eqnarray}
   w_{i\to f}^\mathrm{Al}&=&\frac{e^2}{\pi\hbar \omega^2}
   \int\limits_{(4\pi)} {\rm d}\Omega_{\bm{k}}
   \left| \bm{j}_{fi}(\bm{k})\bm\cdot \bm{k}\right|^2\left|
   \frac{\partial \varepsilon_\mathrm{l}(\omega,k)}
   {\partial k}\right|^{-1}
   _{~k=k_\mathrm{l}(\omega)},
\label{eq:wif_lem}\\
   w_{i\to f}^\mathrm{Atr}&=&\frac{e^2}{\pi\hbar}
   \int\limits_{(4\pi)} {\rm d}\Omega_{\bm{k}}
   \left| \bm{j}_{fi}(\bm{k}) \bm\times \bm{k}\right|^2\left|\frac{\partial
   \left[\omega^2\varepsilon_\mathrm{tr}(\omega,k)-k^2c^2\right]}
   {\partial k}\right|^{-1}
   _{~k=k_\mathrm{tr}(\omega)}.
\label{eq:wif_trem}
\end{eqnarray}
In principle, there can be several poles for one $\omega$; then one
should sum over the poles in these equations.

The processes Bl and Btr in Eq.~(\ref{eq:wif}) involve virtual
plasmons. These processes are allowed if the dielectric functions
have imaginary parts for some values of $k$ at a given  $\omega$.
Nonvanishing imaginary parts of dielectric functions ensure that the
plasma can directly absorb electromagnetic fluctuations induced by
the emitter. From Eq.~(\ref{eq:wif}) one has \cite{Klim84}
\begin{eqnarray}
   w_{i\to f}^\mathrm{Bl}&=&\frac{e^2}{\pi^2\hbar\omega^2} \int\limits_0^\infty
   {\rm d} k\int\limits_{(4\pi)} {\rm d}\Omega_{\bm k}\,
   \frac{\left|\bm{j}_{fi}(\bm{k}) \bm\cdot \bm{k} \right|^2
   \Im \varepsilon_\mathrm{l}(\omega,k)}
   {\left|\varepsilon_\mathrm{l}(\omega,k)\right|^2},
\label{eq:wif_labs}\\
   w_{i\to f}^\mathrm{Btr}&=&\frac{e^2\omega^2}{\pi^2\hbar} \int\limits_0^\infty
   {\rm d} k\int\limits_{(4\pi)} {\rm d}\Omega_{\bm k}\,
   \frac{\left|\bm{j}_{fi}(\bm{k}) \bm\times \bm{k}\right|^2  \Im
   \varepsilon_\mathrm{tr}(\omega,k)}{\left|\omega^2 \varepsilon_\mathrm{tr}
   (\omega,k)-k^2c^2\right|^2}.
\label{eq:wif_trabs}
\end{eqnarray}
Virtual plasmons do not obey any specific dispersion relation and
can have a wide spectrum of wavenumbers $k$ for a given $\omega$.
Equations (\ref{eq:wif_labs}) and (\ref{eq:wif_trabs}) describe the
effects which are in common with collisional broadening of spectral
lines (and associated enhancement of radiative transition rates) in
atomic physics \cite{Klim84}.

Equations (\ref{eq:wif_lem})--(\ref{eq:wif_trabs}) can be used to
study radiative transition rates of a relativistic emitter in any
uniform and isotropic dispersive medium (we set $T\to0$ and
disregard thus induced transitions).


In vacuum, where
$\varepsilon_\mathrm{l}=\varepsilon_\mathrm{tr}\equiv 1$, only one
channel survives out of the four. It is Atr -- the emission of real
transverse plasmons, and these plasmons become identical to ordinary
photons. Then Eq.~(\ref{eq:wif_trem}) reduces to the well known
expression \cite{landau4eng}
\begin{equation}
    w_{i\to f}^\mathrm{Atr}\equiv w_{i\to f}^\mathrm{vac}= \frac{e^2 \omega}
    {2\pi  \hbar c^3} \int\limits_{(4\pi)} {\rm d}\Omega_{\bm{k}}
    \left|\bm{j}_{fi}(\bm{k})\bm\times \bm{\hat{k}} \right|^2,
\end{equation}
where $\bm{\hat{k}}=\bm{k}/k$.

Let us stress that the existence of the four radiative decay
channels and general expressions for the partial decay rates have
been known long ago (e.g., Refs.\ \cite{Klim84,OiKl84}). However,
this formalism has been mostly applied to radiative  transitions
in a non-degenerate plasma, where radiation frequencies are
typically much higher than $\omega_p$. In this case the exchange
of virtual plasmons is usually unimportant, and the authors
focused on the emission of real plasmons that was not strongly
affected by the plasma environment. We will apply the above
formalism to analyze dense degenerate electron stellar matter
where the plasma effects are pronounced much stronger.

\section{Plasma environment of degenerate electrons}
\label{S:diel_funct}

Let us study plasma effects in a strongly degenerate (zero
temperature) ideal electron gas of any degree of relativity in the
absence of a magnetic field. We will comment on the effects of
finite temperature, ion plasma polarization and magnetic fields in
Sec.~\ref{S:dissc}. We employ the collisionless dielectric functions
$\varepsilon_\mathrm{tr}(\omega,k)$ and
$\varepsilon_\mathrm{l}(\omega,k)$ of a relativistic
 electron gas at $T=0$ derived by Jancovici \cite{janc62} in
the random phase approximation. We do not present his cumbersome
expressions here (note that they should be corrected \cite{kg07} at
certain values of $\omega$ and $k$) but discuss their main
properties relevant to our study.

The most important quantity is the electron plasma frequency,
\begin{equation}
      \omega_p= \sqrt{4 \pi e^2 n_e /m_e^\ast},
\label{eq:omega_p}
\end{equation}
where $n_e$ is the electron number density, $m_e^\ast=\mu_e/c^2$ is
the effective electron mass on the Fermi surface,
$\mu_e=\sqrt{m_e^2c^4+c^2 p_F^2}$ the electron chemical potential
(electron rest-mass energy included), $p_F=\hbar (3 \pi^2
n_e)^{1/3}$ being the electron Fermi momentum. We also introduce the
electron Fermi velocity $v_F=p_F/m_e^\ast$.

The functions $\varepsilon_\mathrm{tr}(\omega,k)$ and
$\varepsilon_\mathrm{l}(\omega,k)$ are generally complex. Their real
parts describe plasma effects on the propagation of electromagnetic
fluctuations, while their imaginary parts describe dissipation of
such fluctuations. Under typical parameters in dense stellar matter
for the processes of our study (Sec.~\ref{S:intro}), the main source
of dissipation is provided by the Cherenkov-type absorption at
$\omega \leq k v_F$ (e.g., Ref.~\cite{abr84eng}). At $T=0$ the
dissipation switches on abruptly in this domain; it is absent
whenever  $\omega
> k v_F$.
Thus the integration over $k$ in Eqs.\ (\ref{eq:wif_labs}) and
(\ref{eq:wif_trabs}) can be  truncated at $k=\omega/v_F$.
Furthermore, in dense stellar environment it is reasonable to assume
that radiative transition energies are not too large, $\hbar \omega
\ll v_F p_F$, and $\hbar \omega_p \ll v_F p_F$. This smallness of
$\hbar \omega$ and $\hbar \omega_p$ with respect to typical electron
energies greatly simplifies the consideration.

At $\omega \sim \omega_p$ and $k \ll \omega/v_F$ the dissipation
effect is absent, and the dielectric functions take the form
\begin{eqnarray}
     \varepsilon_\mathrm{l}(\omega,k)&\approx& 1-\frac{\omega_p^2}{\omega^2}
     \left(1+\frac{3}{5}\;\frac{k^2v_F^2}{\omega^2}\right),
\label{eq:epl}\\
     \varepsilon_\mathrm{tr}(\omega,k)&\approx& 1-\omega_p^2/\omega^2.
\label{eq:eptr}
\end{eqnarray}

From Eqs.~(\ref{eq:poles}) and (\ref{eq:eptr}) we immediately obtain
the dispersion relation for the transverse waves
\begin{equation}
     \omega^2_\mathrm{tr} = \omega_p^2 + c^2 k^2.
\label{eq:disptr}
\end{equation}
This is a good approximation for all $k$. It corresponds to two
transverse plasma modes with different polarizations, but the same
dispersion relation. The wave frequency satisfies the inequality
$\omega_\mathrm{tr}> k v_F$ at any $k$. Therefore, these waves
undergo no collisionless damping. One has $\omega_\mathrm{tr} \to
\omega_p$ as $k \to 0$; and $\omega_\mathrm{tr} \approx kc$ as $k
\gg \omega_p/c$. In the latter case these waves turn into ordinary
photons which are almost unaffected by the plasma environment.

From Eqs.~(\ref{eq:poles}) and (\ref{eq:epl}) one can derive the
dispersion relation for the longitudinal (electron Langmuir)
plasma waves \cite{abr84eng},
\begin{equation}
     \omega^2_\mathrm{l} = \omega_p^2 + {3 \over 5} \, v_F^2 k^2.
\label{eq:displ}
\end{equation}
This equation is valid at $k \ll \omega_p/v_F$, when
$\omega_\mathrm{l}$ is only slightly higher than $\omega_p$, and the
collisionless damping is absent. At higher $k$ the dispersion
equation must be solved numerically. The solution shows that at some
$\omega_\mathrm{l}$ ($\sim \omega_p$) the derivative $\partial
\varepsilon_{\rm l}(\omega,k)/\partial k$ becomes very large. This
means that the transition rate $w_{i \to f}^\mathrm{Al}$ switches
off when the transition frequency $\omega$ exceeds some value (a few
$\omega_p$).

It is important that the frequencies of longitudinal and transverse
plasma waves are always higher than $\omega_p$. This implies that
corresponding transition rates undergo the plasma frequency cutoff,
\begin{equation}
       w_{i \to f}^\mathrm{Atr}=w_{i \to f}^\mathrm{Al}=0
       \quad \mathrm{at}~~\omega \leq \omega_p.
\label{eq:cutoff}
\end{equation}

Finally, let us outline absorption properties of degenerate electron
plasma. For typical conditions in dense stellar matter, there are
two domains \cite{janc62}  in the $(\omega,k)$-plane, where the
imaginary parts of the longitudinal and transverse dielectric
functions are non-zero. The first domain is given by the inequality
  $\hbar\omega< \mu-E_{p_F-\hbar k} $ at $\hbar k<2 p_F$;
the second domain is determined by $|E_{p_F-\hbar
k}-\mu|<\hbar\omega< E_{p_F+\hbar k}-\mu$ at any $k$, with $E_{p} =
\sqrt{c^2p^2+ m_e^2c^4}$. The analytic expressions for the imaginary
parts of the dielectric functions in these two domains are
different. Although we have used exact expressions in computations,
we notice that in the ultrarelativistic gas it is sufficient to
consider the only one domain $\omega/v_F\lesssim k\lesssim
2p_F/\hbar$, where, to a good approximation,
\begin{eqnarray}
   \Im\varepsilon_\mathrm{l}&=&\frac{3\pi\omega_p^2}{2\omega^2 x^3}
   \left(1-\frac{x^2 \hbar^2\omega^2}{4p_F^2 c^2} \right),
\label{eq:im_el}\\
   \Im\varepsilon_\mathrm{tr}&=&\frac{3\pi\omega_p^2}
   {4\omega^2 x^3} \left[ x^2-1
   +\frac{x^2 \hbar^2\omega^2}{4p_F^2 c^2}
   \left({x^2c^2\over v_F^2}-1\right)\right],
\label{eq:im_etr}
\end{eqnarray}
with $x=kv_F/\omega$.

\section{Electric dipole transitions}
\label{S:el_dip}

Let us calculate the transition rate $w_{i \to f}$ in the electric
dipole approximation (E1). The approximation is valid at $ka \ll
1$, where $a$ is a typical size of the emitter, and $k$ a typical
plasmon wavenumber. In this case, the transition current is
independent of $\bm{k}$, $\bm{j}_{fi}(\bm{k})
\approx\bm{j}_{fi}({\bm 0})$. The angular integration gives
\begin{eqnarray}
   \int\limits_{(4\pi)} {\rm d}\Omega_{\bm k}
   \left| \bm{j}_{fi}(\bm 0)\bm\cdot \bm{k}\right|^2&=&
   \frac{4\pi}{3}\left|{\bm j}_{fi}({\bm 0})\right|^2 k^2,\\
   \int\limits_{(4\pi)} {\rm d}\Omega_{\bm k}
   \left| \bm{j}_{fi}(\bm 0) \bm\times \bm{k} \right|^2&=&\frac{8\pi}{3}
   \left|{\bm j}_{fi}({\bm 0})\right|^2 k^2.
\end{eqnarray}
Then the in-vacuum transition rate (associated with the emission
of ordinary photons) is given by the standard expression
\cite{landau4eng}
\begin{equation}
   w_{i\to f}^\mathrm{vac}=\frac{4e^2 \omega
   \left|\bm{j}_{fi}(\bm 0)\right|^2}{3\hbar c^3}=
   \frac{4e^2 \omega^3}{3\hbar c^3} \left|\bm{r}_{fi}\right|^2,
\end{equation}
where we have used the relation $\bm{j}_{fi}(\bm 0)\approx -i\omega
\bm{r}_{fi}$, $\bm{r}_{fi}$ being the position-vector  matrix
element.

It is convenient to rewrite Eq.~(\ref{eq:wif_tot}) as
\begin{equation}
\label{eq:wif_R}
   w_{i\to f}=w_{i\to f}^\mathrm{vac} R=w_{i\to f}^\mathrm{vac}
   \left(R_\mathrm{Al}+R_\mathrm{Bl}+R_\mathrm{Atr}+R_\mathrm{Btr}\right),
\end{equation}
where $R_\mathrm{Al}$, $R_\mathrm{Bl}$,
$R_\mathrm{Atr}$, and $R_\mathrm{Btr}$ are the
factors, which describe the plasma effects on the transition rates
in the four channels (Table \ref{tab1}),
and $R$ is the cumulative factor. The partial
factors are given by
\begin{eqnarray}
   R_\mathrm{Al}&=& \frac{c^3}{\omega^3}\,
   \left.
   J_\mathrm{l}(k) k^2 \, \left|\frac{\partial \varepsilon_\mathrm{l}(\omega,k)}
   {\partial k}\right|^{-1}
   \right|
   _{~k=k_\mathrm{l}(\omega)},
\label{eq:Rlem}\\
   R_\mathrm{Atr}&=&\frac{2c^3}{\omega}\,
   \left.
   J_\mathrm{tr}(k)k^2\left|\frac{\partial
   \left[\omega^2\varepsilon_\mathrm{tr}(\omega,k)-k^2c^2\right]}
   {\partial k}\right|^{-1}
   \right|
   _{~k=k_\mathrm{tr}(\omega)},
\label{eq:Rtrem}
\\
  R_\mathrm{Bl}&=&\frac{c^3}{\pi \omega^3} \int\limits_0^\infty
  {\rm d} k \, \frac{J_\mathrm{l}(k)k^2
  \Im \varepsilon_\mathrm{l}(\omega,k)}
  {\left|\varepsilon_\mathrm{l}(\omega,k)\right|^2},
\label{eq:Rlabs} \\
  R_\mathrm{Btr}&=&\frac{2c^3\omega}{\pi} \int\limits_0^\infty
  {\rm d} k \, \frac{J_\mathrm{tr}(k)k^2
  \Im \varepsilon_\mathrm{tr}(\omega,k)}
  {\left|\omega^2\varepsilon_\mathrm{tr}(\omega,k)-k^2c^2\right|^2}.
\label{eq:Rtrabs}
\end{eqnarray}
The functions $J_\mathrm{l}(k)$ and $J_\mathrm{tr}(k)$ describe
non-dipole corrections to the E1 approximation at large $k$ (at $ka
\gg 1$),
\begin{eqnarray}
   J_\mathrm{l}(k)&=&\frac{3}{4\pi}
   \left|\bm{j}_{fi}(\bm 0)\right|^{-2}\int {\rm d}
   \Omega_{\bm k} \left|\bm{j}_{fi}(\bm k)\bm\cdot \hat{\bm{k}}\right|^2,
\label{eq:Jl}\\
   J_\mathrm{tr}(k)&=&\frac{3}{8\pi}
   \left|\bm{j}_{fi}(\bm 0)\right|^{-2}\int {\rm d}
   \Omega_{\bm k} \left|\bm{j}_{fi}(\bm k)
   \bm\times \hat{\bm{k}}\right|^2.
\label{eq:Jtr}
\end{eqnarray}
For $ka \lesssim 1$, we have $J_\mathrm{l}(k) \to 1$ and
$J_\mathrm{tr}(k) \to 1$.

Equations (\ref{eq:Rlem})--(\ref{eq:Rtrabs}) determine the plasma
corrections to the E1 transition rate. Let us calculate them in a
degenerate electron gas.

The emission of real longitudinal and transverse plasmons occurs at
$k\lesssim \omega/v_F$. For the transitions at frequencies $\omega$
not larger than several $\omega_p$ in a dense degenerate electron
gas, one typically has $k\ll p_F/\hbar$. This means, that one can
use the classical dielectric functions
$\varepsilon_\mathrm{l}(\omega,k)$ and
$\varepsilon_\mathrm{tr}(\omega,k)$ \cite{janc62} for calculating
$R_\mathrm{Al}$ and $R_\mathrm{Atr}$. We have checked, that the use
of the exact (quantum) dielectric functions  has no noticeable
effect on the results. Similarly, because for the emission of real
plasmons we typically have $k\ll 1/a$, we can always neglect the
non-dipole corrections in Eqs.~(\ref{eq:Rlem}) and (\ref{eq:Rtrem})
and set $J_\mathrm{l}(k)=J_\mathrm{tr}(k)=1$.
Because no plasmon emission can occur at $\omega<\omega_p$, the
transition rates $w_{i \to f}^\mathrm{Atr}$ and $w_{i \to
f}^\mathrm{Al}$ are suppressed as $\omega \to \omega_p$. Indeed, at
$\omega \sim \omega_p$ the longitudinal and transverse dielectric
functions are given by Eqs.~(\ref{eq:epl}) and (\ref{eq:eptr}).
Using then Eqs.~(\ref{eq:Rlem}) and (\ref{eq:Rtrem}), we find
\begin{eqnarray}
   R_\mathrm{Al}&=&\frac{1}{2}\left(\frac{5}{3}\right)^{3/2}
   \left(\frac{c}{v_F}\right)^3
   \sqrt{1-\frac{\omega_p^2}{\omega^2}}
   \quad \mathrm{at~} 0 \leq \omega - \omega_p \ll \omega_p,
\label{eq:Rlem_appr}\\
   R_\mathrm{Atr}&=&\sqrt{\varepsilon_\mathrm{tr}}
   =\sqrt{1-\frac{\omega_p^2}{\omega^2}}
   \quad \mathrm{at~} \omega > \omega_p.
\label{eq:Rtrem_appr}
\end{eqnarray}
Equation~(\ref{eq:Rtrem_appr}) remains a good approximation at all
$\omega> \omega_p$; in the limit of $\omega \gg \omega_p$ the factor
$R_\mathrm{Atr}$ tends to its in-vacuum value, $R_\mathrm{Atr}=1$.
In contrast, Eq.~(\ref{eq:Rlem_appr}) is valid only for $\omega$
close to $\omega_p$. For higher
$\omega$, the factor $R_\mathrm{Al}$ is strongly suppressed by the
$k$-derivative of the longitudinal dielectric function in
Eq.~(\ref{eq:Rlem}) (no longitudinal plasmons can propagate at
high frequencies, Sec.\ \ref{S:diel_funct}).

The  transition rate at $\omega\lesssim \omega_p$ arises from the
virtual-plasmon channels B, Eqs.~(\ref{eq:Rlabs}) and
(\ref{eq:Rtrabs}). First of all, consider the factor
$R_\mathrm{Bl}$, Eq.~(\ref{eq:Rlabs}). Substituting the
Eq.~(\ref{eq:im_el}) into (\ref{eq:Rlabs}), we obtain
\begin{equation}
  R_\mathrm{Bl}=\frac{3}{2}\left(\frac{c}{v_F}\right)^3
  \left(\frac{\omega_p}{\omega}\right)^2
  \int\limits_1^{x_\mathrm{max}} \frac{{\rm d} x}{x}
  \left(1-\frac{v_F^2}{c^2}\,\frac{x^2}{x_\mathrm{max}^2}\right)
  \frac{J_\mathrm{l}(x\omega/v_F)}
  {\left|\varepsilon_\mathrm{l}(\omega,x\omega/v_F)\right|^2},
\label{eq:Rlabs_app}
\end{equation}
where $x_\mathrm{max}=2p_Fv_F/(\hbar\omega)$. The integrand has a
logarithmic singularity which is avoided owing to a natural
integration cutoff at $x=x_\mathrm{max}$. Therefore, the integral
has the meaning of a Coulomb logarithm. According to
Eq.~(\ref{eq:Rlabs_app}), we can neglect the non-dipole corrections
and set $J_\mathrm{l}(x\omega/v_F)=1$ provided $2p_F\lesssim
\hbar/a$. If so, $R_\mathrm{Bl}$ depends only on plasma
characteristics (and does not depend on the transition properties of
the emitter); in this sense, $R_\mathrm{Bl}$ becomes universal. In
the opposite case of $2p_F \gg \hbar/a$, the function
$J_\mathrm{l}(x\omega/v_F)$ suppresses the integrand in
Eq.~(\ref{eq:Rlabs_app}) at $k\gg 1/a$. Then the cutoff of the
Coulomb logarithm occurs at smaller
$x=v_F/(a\omega)<x_\mathrm{max}$, and the universal factor,
calculated from Eq.~(\ref{eq:Rlabs_app}) with
$J_\mathrm{l}(x\omega/v_F)=1$, gives the upper limit of
$R_\mathrm{Bl}$.

Let us study the universal regime of $2p_F\lesssim \hbar/a$ and
consider the behavior of $R_\mathrm{Bl}$ at $\omega\ll \omega_p$.
In this case we can use the static longitudinal dielectric
function in the denominators of Eq.~(\ref{eq:Rlabs}) or
(\ref{eq:Rlabs_app}). The asymptotic behavior of $R_\mathrm{Bl}$
is
\begin{equation}
   R_\mathrm{Bl}\propto \left( \omega_p / \omega\right)^2
   \quad \mathrm{at}~~\omega \ll \omega_p ,
\end{equation}
implying a strong enhancement of the transition rate over the
in-vacuum rate [although the emission of real plasmons is forbidden,
Eq.~(\ref{eq:cutoff})].

The consideration of the factor $R_\mathrm{Btr}$ is similar. First
of all, substituting Eq.~(\ref{eq:im_etr}) into
Eq.~(\ref{eq:Rtrabs}) we conclude that there is no logarithmic
divergency for the transverse channel. This is because of the extra
$k^2c^2$ term in the denominator of Eq.~(\ref{eq:Rtrabs}). An
analysis shows that the main contribution to $R_\mathrm{Btr}$ comes
from intermediate values of $x$ (whereas the main contribution to
$R_\mathrm{Bl}$ comes from large $x$). As a result, non-dipole
corrections to $R_\mathrm{Bl}$ are much less important than to
$R_\mathrm{Btr}$, and $J_\mathrm{tr}(x\omega /v_F)=1$ is a much
better approximation than $J_\mathrm{l}(x\omega/v_F)=1$. The
asymptotic behavior of $R_\mathrm{Btr}$ is
\begin{equation}
   R_\mathrm{Btr}\propto
   \left( \omega_p / \omega \right)^{2/3}
   \quad \mathrm{at}~~\omega \ll \omega_p .
\end{equation}
Thus, the transitions through the transverse channel are less
efficient than those through the longitudinal channel.

Finally, we have set  $J_\mathrm{l}=J_\mathrm{tr}=1$ and calculated
the total plasma enhancement factor $R$ with the precise dielectric
functions \cite{janc62}. In the case of ultrarelativistic degenerate
electrons ($v_F=c$) the factor $R$ depends on the only one argument
$u=\omega/\omega_p$. In the interval $0.01\leq u \leq 20$ the
numerical results can be fitted by the expression
\begin{equation}
\label{eq:fitE1}
    R=\frac{3.0316}{u^2}\, \left(1+0.13\,u\right)+\Theta(u-1)\,
    \frac{29.3\;(u-1)+(u-1)^3}{0.93+35.6\;(u-1)+(u-1)^3},
\end{equation}
where $\Theta(x)$ is the Heaviside step-function.
The maximum fit error $\approx 1.6\%$ takes place at $u=1.34$. In
the limit of $u\ll 1$ the plasma factor $R$ is dominated by
$R_\mathrm{Bl}$; in the opposite limit of $u \gg 1$ it is dominated by
$R_\mathrm{Atr}$.

\section{Electric quadrupole and magnetic dipole transitions}
\label{S:quadr}

Now consider the case in which the electric dipole transition $i \to
f$ of the emitter is forbidden but the electric quadrupole (E2) or
magnetic dipole (M1) transition is allowed.

Because the E1 transition is forbidden, we have
\begin{equation}
   \bm{j}_{fi}(\bm 0)=\int {\rm d}V \bm{j}_{fi}(\bm{r})=0.
\end{equation}
Multipole transitions are given by next terms in the expansion of
the transition current over $ka$. For the E2 and M1 transitions, the
transition current can be written as
\begin{equation}
     \bm{j}_{fi}(\bm{k})=\int {\rm d} V\, \bm{j}_{fi}(\bm{r})
     \exp(-i\bm{k}\bm{r}) \approx -i\int {\rm d} V \, \bm{j}_{fi}(\bm{r})
     \, (\bm{k}\bm{\cdot}  \bm{r}).
\end{equation}
Using the standard expansion
over spherical vectors $\bm{Y}^L_{JM}$
\cite{Var88}, we obtain
\begin{equation}
\label{eq:jfi_mult}
   \bm{j}_{fi}(\bm{k})=-\frac{4\pi ik}{3} \,\sum_{J=0}^2 \sum_{M=-J}^{J}
   \bm{Y}^1_{JM}(\bm{\hat{k}})\int {\rm d} V\, r\,
   \bm{j}_{fi}(\bm{r})\bm{\cdot} \bm{Y}^{1*}_{JM}(\hat{\bm{r}}),
\end{equation}
where $\hat{\bm{r}}=\bm{r}/r$.
The terms
with different $J$ correspond to  different
transition types.

The term with $J=2$ refers to an E2 transition. Indeed, a rank 2
spherical vector can be presented as \cite{Var88}
\begin{equation}
    \bm{Y}^1_{2M}(\hat{\bm{r}})=
    \frac{1}{\sqrt{10} \, r}\; \grad \left(r^2
    Y_{2M}(\hat{\bm{r}})\right),
\end{equation}
where $Y_{2M}(\hat{\bm{r}})$ is a spherical function.
Then one can rearrange the integral term in Eq.~(\ref{eq:jfi_mult})
as
\begin{equation}
   \int {\rm d} V \,
   r \, \bm{j}_{fi}(\bm{r})\bm{\cdot} \bm{Y}^{1*}_{2M}(\hat{\bm{r}}) =
   \frac{i\omega}{\sqrt{10}}\int {\rm d}V\, \rho_{fi}(\bm{r})\,
   r^2 Y^*_{2M}(\hat{\bm{r}})
   =(-1)^M\frac{i\omega }{\sqrt{8\pi}}\, Q^{(e)}_{2-M},
\end{equation}
where $Q_{2-M}^{(e)}$ is an electric quadrupole component
\cite{landau4eng} and $\rho_{fi}(\bm{r})$ is the matrix element of
the density operator, that is related to the transition current
through the continuity equation
\begin{equation}
   i \omega \rho_{fi}(\bm{r}) + \divv\; \bm{j}_{fi}(\bm{r})=0.
\end{equation}
%

The $J=1$ term in Eq.~(\ref{eq:jfi_mult}) corresponds to an M1
transition. Using the relation
\begin{equation}
   \bm{Y}^{1}_{1M}(\hat{\bm{r}})=-\frac{i}{\sqrt{2}}\,
   \bm{r}\bm{\times}\grad\; Y_{1M}(\hat{\bm{r}}),
\end{equation}
one finds \cite{landau4eng}
\begin{eqnarray}
  \int{\rm d} V r \,
   \bm{j}_{fi}(\bm{r})\bm{\cdot} \bm{Y}^{1*}_{1M}(\hat{\bm{r}})&=& \frac{i}{\sqrt{2}}
   \int {\rm d} V\, r
   \bm{j}_{fi}(\bm{r})\bm{\cdot} \left[\bm{r} \bm{\times} \grad\; {Y}^*_{1M}
   (\hat{\bm{r}})\right]
\nonumber   \\
  & =& -i(-1)^M\sqrt{\frac{3}{2\pi}}\;
   Q^{(m)}_{1-M},
\end{eqnarray}
where $Q^{(m)}_{1-M}$ is a component of the magnetic dipole moment.

Finally, the $J=0$ term in Eq.~(\ref{eq:jfi_mult}) is non-standard.
It is absent in vacuum, but appears in plasma.
Rearranging the spatial integration in Eq.~(\ref{eq:jfi_mult}) in
the same way, as for E2 and M1 transitions, we obtain
\begin{equation}
   \int {\rm d} V \, r \,
   \bm{j}_{fi}(\bm{r})\bm{\cdot} \bm{Y}^{1*}_{00}(\hat{\bm{r}})
   =-\frac{i\omega}{2\sqrt{4\pi}}
   \int {\rm d} V\, \rho_{fi}(\bm{r})\, r^2=-\frac{i\omega}{2\sqrt{4\pi}}\, Q_2,
\end{equation}
where
%
\begin{equation}
   Q_2\equiv\int {\rm d} V\, \rho_{fi}(\bm{r})\, r^2.
\end{equation}

In order to separate the contributions from the above terms to the
longitudinal and transverse transition channels let us split the
$\bm{\hat{k}}$-dependent spherical vectors into components
longitudinal and transverse to $\bm{\hat{k}}$. The $J=2$ term
contains both, transverse and longitudinal, components \cite{Var88}:
\begin{equation}
   \bm{Y}^1_{2M}(\bm{\hat{k}})=
   \sqrt{\frac{3}{5}} \, \bm{Y}^{(1)}_{2M}(\bm{\hat{k}})
   +\sqrt{\frac{2}{5}} \, \bm{Y}^{(-1)}_{2M}(\bm{\hat{k}}),
\end{equation}
where $\bm{Y}^{(-1)}_{2M}(\bm{\hat{k}})$ is the longitudinal
spherical vector and $\bm{Y}^{(1)}_{2M}(\bm{\hat{k}})$ is the
transverse electric-type spherical vector. The $J=1$ term in
(\ref{eq:jfi_mult}) contains only the transverse
component,
\begin{equation}
   \bm{Y}^{1}_{1M}(\bm{\hat{k}})=\bm{Y}^{(0)}_{1M}(\bm{\hat{k}}),
\end{equation}
%
$\bm{Y}^{(0)}_{1M}(\hat{\bm{r}})$
being the transverse magnetic vector. The $J=0$ term in
(\ref{eq:jfi_mult}) contains only the longitudinal vector,
\begin{equation}
    \bm{Y}^1_{00}(\bm{\hat{k}})=-\bm{Y}^{(-1)}_{00}(\bm{\hat{k}}).
\end{equation}

Now we can calculate the transition rate from Eq.~(\ref{eq:wif}).
Performing angular integration, we find for the transverse channel:
\begin{equation}
   \int {\rm d} \Omega_{\bm k} \,
   \left| \bm{k} \bm\times \bm{j}_{fi}(\bm k) \right|^2
   =\frac{8\pi}{3}k^4 \,
   \left[\sum_M \left|Q^{(m)}_{1-M}\right|^2+\frac{\omega^2}{20}\; \sum_M
   \left|Q^{(e)}_{2-M}\right|^2\right].
\end{equation}
The two terms in the square brackets correspond to the M1 and E2
transitions, respectively. No contribution from the $J=0$ is
present, because the $J=0$ term is purely longitudinal. In vacuum,
only the transverse channel contributes to the transition
rate. Then from Eq.~(\ref{eq:wif_trem}) we recover the well-known
expression
\begin{equation}
     w_{i\to f}^\mathrm{vac}=w_{i\to f}^\mathrm{Atr}
     =\frac{4\omega^3}{3\hbar c^5}
     \left[\sum_M
    \left|Q^{(m)}_{1-M}\right|^2+\frac{\omega^2}{20}\; \sum_M
     \left|Q^{(e)}_{2-M}\right|^2\right]\equiv
    w_\mathrm{vac}^\mathrm{M1}+w_\mathrm{vac}^\mathrm{E2}.
\end{equation}

The angular integration of the longitudinal part of
Eq.~(\ref{eq:wif}) gives
\begin{equation}
\label{eq:int_mul_l}
   \int {\rm d} \Omega_{\bm{k}} \,
   \left|\bm{k} \bm\cdot \bm{j}_{fi}(\bm k)\right|^2
   =\frac{4\pi}{45}\,k^4\omega^2
   \left[\sum_M
   \left|Q^{(e)}_{2-M}\right|^2+\frac{5}{4}\left|Q_2\right|^2\right].
\end{equation}
The first term in the square brackets corresponds to the E2
transition, while the second term refers to a different, purely
longitudinal transition \cite{OiKl84}. The latter transition is not
forbidden by the standard selection rules and should be kept in line
with the E2 transition (at $M=0$). Note, that this term can be
presented as a trace of the quadrupole moment tensor of the emitter,
$Q_2={\rm Tr}\{ Q_{\alpha\beta}\}$, if this tensor is defined in a
non-standard (not irreducible) form as $Q_{\alpha\beta}=\int {\rm d}
V\, \rho_{fi}(\bm{r})\; x_\alpha x_\beta$. If so, the $Q_2$ term can
be regarded as an additional contribution to the quadrupole
transition \cite{OiKl84}; nevertheless it can be presented even for
a spherical emitter. Let us add, that there is no M1 transition in
Eq.~(\ref{eq:int_mul_l}) -- it is forbidden because the M1
transition current is purely transverse.

The plasma effects on the transition rate can be described by
introducing the plasma factors $R$ in accordance with
Eq.~(\ref{eq:wif_R}). We obtain
\begin{eqnarray}
  w^\mathrm{E2}_{i\to f}&=&w^\mathrm{E2}_\mathrm{vac}R_\mathrm{E2}
  =w^\mathrm{E2}_\mathrm{vac} \left(R^{(2)}_\mathrm{Al}+R^{(2)}_\mathrm{Atr}
  +R^{(2)}_\mathrm{Bl}+R^{(2)}_\mathrm{Btr}\right),
\label{eq:wE2_R}\\
  w^\mathrm{M1}_{i\to f}&=&w^\mathrm{M1}_\mathrm{vac}R_\mathrm{M1}
  =w^\mathrm{M1}_\mathrm{vac}
  \left(R^{(2)}_\mathrm{Atr}+R^{(2)}_\mathrm{Btr}\right),
\label{eq:wM1_R}
\end{eqnarray}
where
\begin{eqnarray}
  R_\mathrm{Al}^{(2)}&=& \frac{4 c^5k^4}{3 \omega^5}
  \, \left|\frac{\partial \varepsilon_\mathrm{l}(\omega,k)}
  {\partial k}\right|^{-1}
  _{~k=k_\mathrm{l}(\omega)},
\label{eq:Rlem_quadr}\\
  R_\mathrm{Atr}^{(2)}&=&\frac{2 c^5 k^4}{\omega^3}\,
  \left|\frac{\partial \left[\omega^2\varepsilon_\mathrm{tr}(\omega,k)
  -k^2c^2\right]}
  {\partial k}\right|^{-1}
  _{~k=k_\mathrm{tr}(\omega)},
\label{eq:Rtrem_quadr}\\
  R_\mathrm{Bl}^{(2)}&=&\frac{4 c^5}{3 \pi \omega^5}
  \int\limits_0^\infty {\rm d} k \, \frac{k^4
  \Im \varepsilon_\mathrm{l}(\omega,k)}
  {\left|\varepsilon_\mathrm{l}(\omega,k)\right|^2},
\label{eq:Rlabs_quadr} \\
  R_\mathrm{Btr}^{(2)}&=&\frac{2c^5}{\pi \omega} \int\limits_0^\infty
  {\rm d} k \, \frac{k^4 \Im
  \varepsilon_\mathrm{tr}(\omega,k)}{\left|\omega^2\varepsilon_\mathrm{tr}
  (\omega,k)-k^2c^2\right|^2};
\label{eq:Rtrabs_quadr}
\end{eqnarray}
the upperscript $(2)$ marks the second-order multipole expansion.
Equations~(\ref{eq:Rlem_quadr})--(\ref{eq:Rtrabs_quadr}) differ from
Eqs.~(\ref{eq:Rlem})--(\ref{eq:Rtrabs}) by powers of $k$ in the
numerators ($k^4$ instead of the $k^2$) and by pre-factors.
%
Moreover, the total
transition rate in the second-order multipole expansion contains an
additional term \cite{OiKl84} in the longitudinal channel,
\begin{equation}
   w_{i\to f}= w^\mathrm{M1}_{i\to f}
   +w^\mathrm{E2}_{i\to f}+w^\mathrm{L}_{i\to f},
\end{equation}
where
\begin{equation}
   w^\mathrm{L}_{i\to f}
   =w^\mathrm{E2}_\mathrm{vac}\frac{5\left|Q_2\right|^2}
   {4\sum\limits_{M}\left|Q^{(e)}_{2-M}\right|^2}
   \left(R^{(2)}_\mathrm{Al}+R^{(2)}_\mathrm{Bl}\right).
\label{eq:wL}
\end{equation}
In order to calculate this term one should know transition matrix
elements.

We have calculated and fitted the factors $R_\mathrm{E2}$ and
$R_\mathrm{M1}$ under the same assumptions as the factor $R$ for the
E1 transitions. We considered an ultrarelativistic degenerate
electron gas ($v_F=c$) and employed exact dielectric functions
\cite{janc62}. Again,  $R_\mathrm{E2}$ and $R_\mathrm{M1}$ become
functions of the only one parameter  $u=\omega/\omega_p$ which was
varied in the range $0.01\leq u \leq 20$. Our fit to $R_\mathrm{E2}$
is
\begin{equation}
   R_\mathrm{E2}=1+\frac{607.8}{u^4}\left(1+0.0048 u^2\right),
\end{equation}
with
the maximum fit error of $1.6\%$ at $u=7.65$. Note, that the
function $R^{(2)}_\mathrm{Bl}$ does not deviate from its small-$u$
asymptotic behavior $R^{(2)}_\mathrm{Bl}=607.8\,u^{-4}$ in the
entire fit interval.

The fit to $R_\mathrm{M1}$ is
\begin{equation}
   R_\mathrm{M1}=1+\frac{1}{u^2}\left(4.82-0.7\ln u-0.47\,u\right),
\end{equation}
with the
maximum fit error of $3.8\%$ at $u=0.1$.

\section{Discussion}
\label{S:dissc}

\begin{figure}[th]
\includegraphics[width=0.4\textwidth]{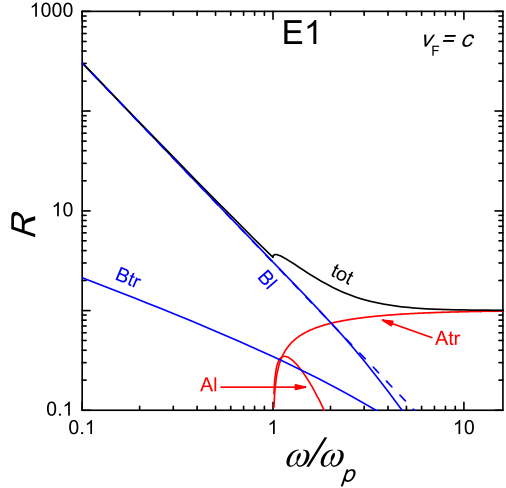}

\caption{(Color online). Different partial contributions to the
total plasma enhancement factor $R$ (curve `\textrm{tot}') as a
function of $\omega/\omega_p$ for E1 transitions at $v_F=c$. Other
solid curves show $R_\mathrm{Btr}$, $R_\mathrm{Bl}$,
$R_\mathrm{Al}$, and $R_\mathrm{Atr}$. The dashed curve is the
low-$\omega$ asymptote $R_\mathrm{Bl}=3.03\, \omega_p^2/\omega^2$.}
\label{fig:E1}
\end{figure}

In Fig.~\ref{fig:E1} we plot various plasma factors for an E1
transition rate as a function of $\omega/\omega_p$ in the
ultrarelativistic strongly degenerate electron plasma ($v_F=c$) at
$k_\mathrm{max}a \ll 1$. The solid line marked `tot' shows the
total plasma enhancement factor $R$. Other solid curves are
partial contributions $R_\mathrm{Btr}$, $R_\mathrm{Bl}$,
$R_\mathrm{Al}$, and $R_\mathrm{Atr}$ given by Eqs.\
(\ref{eq:Rlem})--(\ref{eq:Rtrabs}). The factors $R_\mathrm{Al}$
and $R_\mathrm{Atr}$ vanish at $\omega<\omega_p$ because no real
plasma waves can be emitted under such conditions. At $\omega
\gtrsim 3\omega_p$ the main contribution to the total transition
rate comes from the emission of real transverse plasmons. In the
limit of $\omega \gg \omega_p$ the plasma effects disappear and
$R\approx R_\mathrm{Atr}\to 1$.  Radiative transitions via virtual
longitudinal plasmons always dominate over transitions via virtual
transverse plasmon, $R_\mathrm{Bl}>R_\mathrm{Btr}$.
All transitions at $\omega<\omega_p$ go via the exchange of virtual
plasmons, the transition rate being greatly enhanced in comparison
with its in-vacuum value. The dashed line in Fig.~\ref{fig:E1} shows
the low-$\omega$ asymptote
$R_\mathrm{Bl}=3.03\,\left(\omega_p/\omega\right)^{2}$; it is
accurate at $\omega<\omega_p$, where $R_\mathrm{Bl}$ dominates.


\begin{figure}[t]
\includegraphics[width=0.4\textwidth,bb= 90 40 570 510]{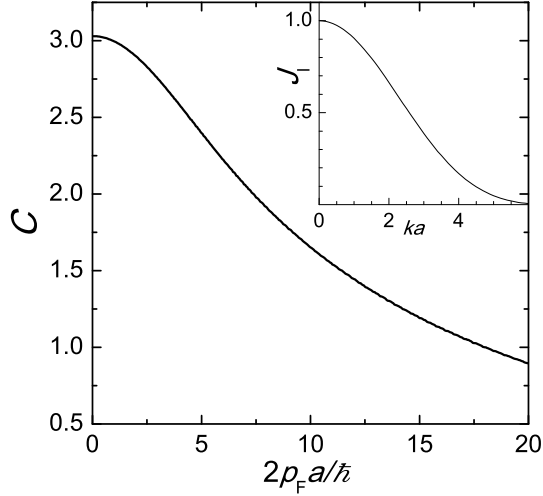}
\caption{
  Factor $C$ versus $k_\mathrm{max}a$
  as an illustration of the effect of non-dipole corrections on
  $R_\mathrm{Bl}$ in Eq.~(\ref{eq:newasy}) for a simplified model of
  radiative deexcitation of atomic nucleus of radius $a$ (see text). The inset
  shows $J_\mathrm{l}(ka)$ for this model.
  }
\label{fig:E1_xmax}
\end{figure}

All quantities, plotted in Fig.~\ref{fig:E1}, are calculated
neglecting non-dipole corrections to the transition current [by
setting $J_\mathrm{l}(k)=J_\mathrm{tr}(k)=1$ that is valid at
$k_\mathrm{max} a = 2 p_F a/\hbar \ll 1$, see Eqs.~(\ref{eq:Jl}) and
(\ref{eq:Jtr})]. The factors $R$ in Fig.~\ref{fig:E1} are universal,
and depend  only on $\omega/\omega_p$ (in the limit of $v_F \to c$).
However, the condition $k_\mathrm{max} a \ll 1$ can be violated.
Such a violation does not significantly affect $R_\mathrm{Al}$,
$R_\mathrm{Atr}$, and $R_\mathrm{Btr}$ but can change
$R_\mathrm{Bl}$ (Sec.~\ref{S:el_dip}). The effect of non-dipole
corrections on $R_\mathrm{Bl}$ is demonstrated in
Fig.~\ref{fig:E1_xmax}. Now the universality is lost and the result
depends on the specific form of the function $J_\mathrm{l}(k)$
(determined by the wave functions of the emitter). For illustration,
we consider a dipole transition from the lowest excited state with
orbital momentum $L=1$ to the ground state ($L=0$) in a spherical
potential well of radius $a$ with infinitely high walls (as a very
rough model of E1 deexcitation of atomic nucleus). The appropriate
function $J_\mathrm{l}(ka)$ is shown in the inset.
At $\omega \ll \omega_p$ we still obtain the asymptote
\begin{equation}
\label{eq:newasy}
  R_\mathrm{Bl}=C(k_\mathrm{max} a)(\omega_p/\omega)^2,
\end{equation}
where $C(k_\mathrm{max}a)$ is now determined by $J_\mathrm{l}(ka)$
(and does not depend of $\omega_p/\omega$). The function
$C(k_\mathrm{max} a)$ takes into account the non-dipole corrections.
It is shown in Fig.~\ref{fig:E1_xmax} for our particular model. The
increase of $k_\mathrm{max}a$ reduces $C(k_\mathrm{max} a)$ with
respect to its purely dipole limit $C(0) \approx 3.03$. A reduction
by a factor of 2 is achieved at $k_\mathrm{max}a \approx 10$.

Thus, non-dipole corrections lower the transition rate, but the rate
remains enhanced over its in-vacuum level by a factor of
$(\omega_p/\omega)^2$.This is because the expression for
$R_\mathrm{Bl}$ (at $\omega<\omega_p$) contains the factor
$\left(\omega_p/\omega\right)^{2}$ which arises from
$\Im\,\varepsilon_\mathrm{l}(\omega,k)$. The integration over $k$ in
Eq.~(\ref{eq:Rlabs_app})  can be carried out using static
 dielectric function $\varepsilon_\mathrm{l}(0,k)$; the function $J_\mathrm{l}(k)$
 specifies only a
numerical prefactor, but does not violate the
$\left(\omega_p/\omega\right)^{2}$ dependence.

\begin{figure}[ht]
\includegraphics[width=0.4\textwidth]{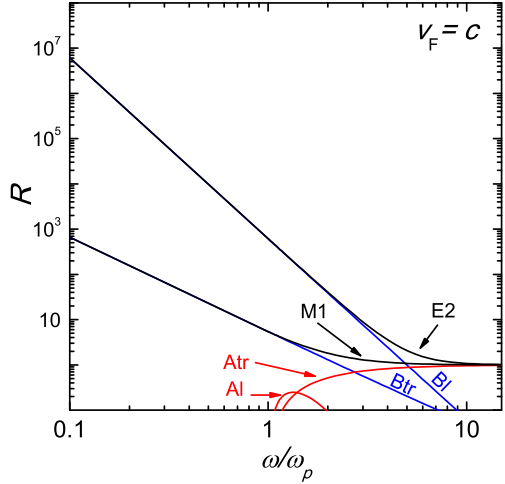}
\caption{(Color online). Plasma enhancement factors $R_\mathrm{E2}$
  and $R_\mathrm{M1}$ 
  versus $\omega/\omega_p$ at $v_F=c$, together with the factors
  $R_\mathrm{Bl}^{(2)}$, $R_\mathrm{Btr}^{(2)}$, $R_\mathrm{Al}^{(2)}$, and
  $R_\mathrm{Atr}^{(2)}$ for separate transition channels. }
\label{fig:Quadr}
\end{figure}

Now let us consider the E2 and M1 radiative transitions. Their
principal features remain the same as for the E1 transitions. In
Fig.~\ref{fig:Quadr} we plot the plasma enhancement factors
$R_\mathrm{E2}$ and $R_\mathrm{M1}$. One can see that the plasma
enhancement of E2 and M1 transitions is stronger than for E1
transitions. However, this is true only if the E1 transition is
forbidden. If not, the effect of higher-order transitions
(particularly, E2 and M1) is included in the functions
$J_\mathrm{l}(k)$ and $J_\mathrm{tr}(k)$ and has already been
discussed above.


Other curves in Fig.~\ref{fig:Quadr}
show partial contributions to the plasma enhancement factors from
different radiative decay channels. The main contribution to E2
transitions comes from $R^{(2)}_\mathrm{Bl}$ (as for E1
transitions). Typically higher values of $R^{(2)}_\mathrm{Bl}$ [with
respect to $R_\mathrm{Bl}$ for the E1 case, see
Eq.~(\ref{eq:Rlabs})], result from the appearance of an additional
$k^2$ in the numerator of Eq.~(\ref{eq:Rlabs_quadr}). This leads to
a stronger $\omega_p/\omega$ dependence in the asymptotic behavior
of $R^{(2)}_\mathrm{Bl}$, in comparison with $R_\mathrm{Bl}$,
$\left(\omega_p/\omega\right)^{4}$ instead of
$\left(\omega_p/\omega\right)^{2}$. Note that the asymptotic
expression
$R^{(2)}_\mathrm{Bl}=607.8\left(\omega_p/\omega\right)^{4}$ remains
an excellent approximation in the entire range of $\omega$ presented
in Fig.~\ref{fig:Quadr}.  Note also, that we have used the first
nonvanishing term in the series expansion of the transition current
over $ka$. Therefore, our results for E2 and M1 transitions are
valid for $2p_F\ll \hbar/a$. In the opposite case, just as for the
E1 transitions, the results will depend on the exact form of the
local transition current $\bm{j}_{fi}(\bm{r})$

We cannot plot the contribution of the additional term, $w_{i\to
f}^\mathrm{L}$ [see Eq.~(\ref{eq:wL})], to the total transition rate
in a similar universal form. If the corresponding moments were
equal, then the transition rate in the longitudinal channel in a
plasma at $\omega \lesssim \omega_p$ would be about twice larger
than the E2 transition rate. For $\omega \gg \omega_p$ the
transition rate $w_{i\to f}^\mathrm{L}$ vanishes.

\begin{figure}[t]
\includegraphics[width=0.4\textwidth,bb= 90 40 570 510]{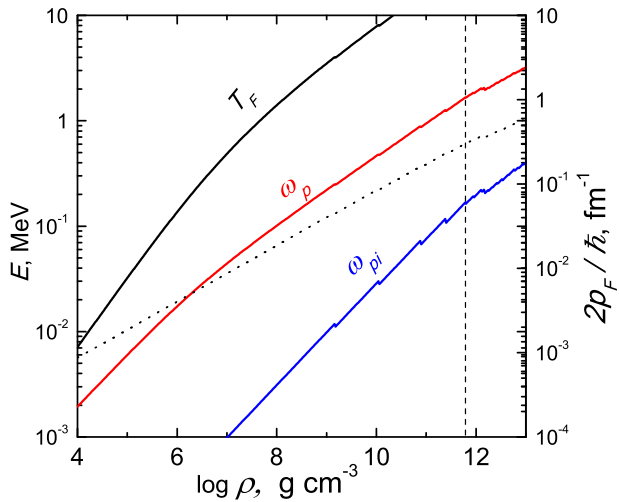}
\caption{
  Energy-density diagram (left vertical scale, solid curves) for dense stellar matter.
  The curves $T_F$, $\omega_p$ and $\omega_{pi}$ show, respectively, the Fermi
  energy of degenerate electrons, electron plasma energy $\hbar
  \omega_p$, and ion plasma energy $\hbar \omega_{pi}$ for an accreting neutron
  star. The short-dashed vertical line positions the neutron drip
  point.
  The dotted line (right vertical scale) shows $k_\mathrm{max}$ versus
  $\rho$ to characterize
  the importance of non-dipole corrections for E1 transitions.
  See text for details.
  }
\label{fig:diag}
\end{figure}

While calculating the $R$-factors, we have used the dielectric
function of the degenerate electron gas and have neglected the ion
contribution. This approximation is expected to be valid for
transition frequencies $\omega$ which are much higher than the ion
plasma frequency $\omega_{pi}$ (see Fig.~\ref{fig:diag}). Because
$\omega_{pi} \ll \omega_p$, the transition frequencies $\omega
\lesssim \omega_{pi}$, at which the ion contribution can be
important, are much lower than the electron plasma frequency. If
necessary, the ion contribution can be studied using similar
approach.

The same plasma effects occur in a magnetized plasma but the
magnetic field complicates the problem. Because of the anisotropy,
introduced into the plasma polarization properties by the magnetic
field, the plasma waves (plasmons) become of mixed type (neither
longitudinal, nor transverse) and have many branches (for instance,
electron cyclotron modes). The properties of the plasmon emission
(channels Al and Atr) of an atom in a rarefied magnetoactive cosmic
plasma were studied, for instance, in Refs.\ \cite{ko74,OiKl84}. The
effect of the magnetic field on the processes with virtual plasmons
seems to be unexplored.

In our analysis, we have employed zero-temperature approximation but
similar effects should be pronounced at finite temperatures.
Moreover, thermal plasma fluctuations, available in this case
\cite{Weisheit88}, can power inverse transitions and excite the
emitter. The efficiency of inverse transitions depends on
temperature and plasma parameters.

The plasma effects are important for studying a number of phenomena
in neutron stars and white dwarfs. These effects are outlined below
using the energy-density diagram for dense stellar matter
(Fig.~\ref{fig:diag}, left vertical scale). The solid lines marked
as $T_F$, $\omega_p$, and $\omega_{pi}$ show the density dependence
of the electron degeneracy energy $k_B T_F$ ($T_F$ being the
electron degeneracy temperature), as well as the electron and ion
plasma energies, $\hbar \omega_p$ and $\hbar \omega_{pi}$. 
For simplicity, we employ the model of accreted neutron star crust
(Table A3 from Ref.\ \cite{hz07}).
The curves $T_F$ and $\omega_p$ are rather insensitive to possible
variations of nuclear composition in an accreting neutron star (and
quite close to the curves for a neutron star whose crust is composed
of the ground-state matter \cite{hpyBOOK}). The curve $\omega_{pi}$
is more sensitive to the composition but is relatively unimportant
for our analysis. Note that the neutron drip occurs at $\rho \approx
6 \times 10^{11}$ g~cm$^{-3}$ in the accreted crust (and at $\rho
\approx 4 \times 10^{11}$ g~cm$^{-3}$ in the ground-state matter).
Typical temperatures in neutron stars and white dwarfs are below
$10^9$~K ($k_BT \lesssim 0.1$ MeV). Degenerate electrons become
relativistic at $\rho \gtrsim 10^6$ g~cm$^{-3}$.

Plasma effects can affect beta captures in dense stellar matter, for
instance, in the crust of an accreting neutron star in a binary
system with a low-mass companion. Such systems manifest themselves
as X-ray transients which demonstrate periods of active accretion
and quiescence \cite{csl97}. Observations show that neutron stars in
X-ray transients remain warm during quiescent periods that is often
explained \cite{bbr98} by deep crustal heating associated with
nuclear transformations \cite{hz90a,gu07,hz07}, particularly, beta
captures, in the accreted matter. When the accreted matter is
gradually compressed by newly accreted material, the density in
local matter elements goes up increasing the Fermi energy of
degenerate electrons. This triggers beta captures with the
appearance of daughter nuclei in ground or excited states. If the
daughter nuclei are born in the excited states, they can de-excite
through radiative transitions \cite{gu07}; the associated energy
release can contribute to the deep crustal heating. The nuclear
composition of the accreted matter can be very different and contain
a wide spectrum of nuclides \cite{gu07}. This means numerous beta
captures involving various nuclei at the densities up to
$10^{11}-10^{12}$ g~cm$^{-3}$ (Fig.\ \ref{fig:diag}). In this case
the electron gas is strongly degenerate and ultrarelativistic, the
electron plasma energy $\hbar \omega_p$ can reach a few MeV and
become larger than transition energies $\hbar \omega$ in some
nuclei. What will happen with these nuclei? A naive answer would be
that they would not decay to lower states because they cannot emit
any electromagnetic quanta at $\omega<\omega_p$. Our results show
quite the opposite. The plasma environment  enhances the decays
through the processes B involving virtual (mostly longitudinal)
plasmons. The dotted line in Fig.~\ref{fig:diag} plots (right
vertical scale) the values of $k_\mathrm{max}$. As follows from the
above discussion (see Fig.~\ref{fig:E1_xmax}), at $k_\mathrm{max}a
\gtrsim 10$ our plasma enhancement factor for E1 transitions starts
to deviate from the universal enhancement (\ref{eq:fitE1}). The
dotted line indicates that, for typical radii $a$ of atomic nuclei,
we have $k_\mathrm{max}a \lesssim 10$ at any $\rho$ in
Fig.~\ref{fig:diag}, so that the enhancement remains universal. Let
us stress, however, that we use a very crude model of E1 transition
in Fig.~\ref{fig:E1_xmax}. We would advise to check the condition
for the breaking of universality (\ref{eq:fitE1}) in specific
situations.

Another example is provided by the reactions involving neutrons (n)
in accreting neutron stars \cite{gkm08}. Specifically, we mean the
reactions (n,$\gamma$) and ($\gamma$,n) (neutron absorption by a
nucleus with the emission of electromagnetic quantum, and an inverse
process). These reactions can occur at densities $10^{11}-10^{12}$
g~cm$^{-3}$ near the neutron drip density in the neutron star crust
(Fig.~\ref{fig:diag}). They can accompany deep nuclear burning of
accreted matter and affect energy release and nuclear
transformations in deep crustal heating process as well as X-ray
superbursts (highly energetic X-ray bursts demonstrated by some
accreting neutron stars). Again, many nuclei can be involved, and
typical energies $\hbar\omega$ of electromagnetic transitions can be
lower than $\hbar \omega_p$. Our results cannot be used directly to
study the neutron reactions, but they can be modified for that
purpose. They demonstrate that the plasma effects cannot suppress
\cite{gkm08} the neutron capture reactions (n,$\gamma$) at
$\omega_p> \omega$. Moreover, we can expect that even at $\omega \ll
\omega_p$, but at not very low temperatures, there will be a
substantial level of fluctuating plasma microfields (associated with
virtual plasmons) to power the inverse reaction ($\gamma$,n).

Finally, the present results can be useful for calculating the
radiative thermal conductivity in a degenerate electron gas. This is
an important problem for outer cores of white dwarfs and outer
envelopes of neutron stars, where the radiative conduction becomes
comparable to the electron one (the latter dominates in the deeper,
strongly degenerate layers of these objects; see, e.g., Ref.\
\cite{pcy97}). With increasing density into the degenerate matter,
the electron plasma frequency becomes comparable to typical
radiative transition frequencies ($\omega \sim k_BT/\hbar$) and then
exceeds them.  Radiative conduction is provided by real
electromagnetic waves (not virtual excitations), which leads to the
plasma cutoff of the radiative thermal conductivity at low
temperatures ($k_B T \ll \hbar \omega_p$, Fig.~\ref{fig:diag}). This
cutoff has been mentioned in the astrophysical literature (e.g.,
\cite{by76}). A general physical theory of radiative transfer in
dispersive media was constructed long ago \cite{harris65}. Several
attempts have been made (e.g., \cite{ao79,kvh94}) to calculate the
radiative thermal conductivity in dense stellar matter with account
for the plasma effects. However, these calculations have neglected
the contribution of longitudinal plasmons which is expected to be
important at $k_B T \lesssim \hbar \omega_p$, especially in the
non-relativistic mildly degenerate electron gas ($T \sim T_F$) where
the radiative thermal conductivity can be comparable with the
electron one.


\section{Conclusions}
\label{S:concl}

We have analyzed the radiative transition rate of an emitter (an
atom or atomic nucleus) immersed in a dense degenerate plasma. Such
a transition goes, generally, through four channels  which involve
real and virtual longitudinal and transverse plasmons (Refs.\
\cite{Klim84,OiKl84}; Table \ref{tab1}).
The emission of real
plasmons is allowed only at radiative transition frequencies
$\omega$ higher than the electron plasma frequency $\omega_p$. The
processes with virtual plasmons operate at any $\omega$.

Our main conclusions are:
\begin{enumerate}

\item The cumulative effect of the plasma is to enhance the
radiative decay rate over the standard radiative decay rate through
the emission of photons in vacuum. In the limit of $\omega \gg
\omega_p$ the plasma enhancement effect disappears.

\item The enhancement becomes especially strong at $\omega \ll
\omega_p$ (where real plasmons cannot exist at all), being mainly
provided by
processes with virtual longitudinal plasmons.

\item The plasma enhancement  takes place for electric
dipole transitions, and for higher-order transitions (such as
electric quadrupole and magnetic dipole one); it is more
pronounced for higher-order transitions.

\item In a strongly degenerate ultrarelativistic electron plasma
the plasma enhancement depends mainly on the parameter
$\omega/\omega_p$. This dependence is calculated and approximated by
analytic expressions for E1, E2 and M2 transitions.

\end{enumerate}

The plasma enhancement effects can strongly modify radiative thermal
conduction in dense stellar matter, kinetics of atomic nuclei in
excited states, emission and absorption of neutrons. Such effects
can be important in degenerate cores of white dwarfs and envelopes
of neutron stars but are almost unexplored.

\begin{acknowledgments}
We are grateful to H.\ Schatz, who drew our attention to the problem
of study, and to D.~A.\ Varshalovich for useful discussions. The
work is supported by the Russian Foundation for Basic Research
(grant 08-02-00837a), by the State Program ``Leading Scientific
Schools of Russian Federation'' (grant NSh 2600.2008.2).
\end{acknowledgments}

\end{document}